\documentclass[a4paper,11pt]{article}

\usepackage{aas_macros}
\usepackage{wrapfig}
\usepackage{jcappub}
\usepackage{amssymb}
\usepackage{amsmath}
\usepackage{color,units}
\usepackage{xspace}
\usepackage{subcaption}
\usepackage{comment}
\usepackage{xcolor}
\usepackage{bm}
\usepackage{dcolumn}
\usepackage{enumitem} 
\usepackage{array}
\usepackage{nicematrix}
\usepackage{hhline}
\usepackage{graphicx}
\usepackage{multirow}
\usepackage[normalem]{ulem}
\usepackage{soul}
\usepackage[math]{cellspace}
\usepackage{nicematrix} 
\usepackage{makecell}
\usepackage{acronym}
\usepackage{caption}
\usepackage{xspace}

\usepackage{hyperref}

\newcommand{\GammaBayes}{\texttt{GammaBayes}\xspace}

\usepackage{tikz}

\begin{document}

\title{\GammaBayes: a Bayesian pipeline for dark matter detection with CTA}

\author[a]{Liam Pinchbeck}
\emailAdd{Liam.Pinchbeck@monash.edu}

\author[a,b]{Eric Thrane}

\author[a]{Csaba Balazs}

\affiliation[a]{School of Physics and Astronomy, Monash University, VIC 3800, Australia}
\affiliation[b]{OzGrav: The ARC Centre of Excellence for Gravitational-Wave Discovery, Clayton, VIC 3800, Australia}

\abstract{
We present \GammaBayes, a Bayesian \texttt{Python} package for dark matter detection with the Cherenkov Telescope Array (CTA).
\GammaBayes takes as input the CTA measurements of gamma rays and a user-specified dark-matter particle model.
It outputs the posterior distribution for parameters of the dark-matter model including the velocity-averaged cross section for dark-matter self interactions $\langle\sigma v\rangle$ and the dark-matter mass $m_\chi$.
It also outputs the Bayesian evidence, which can be used for model selection.
We demonstrate \GammaBayes using {\unit[525]{hours}} of simulated data, corresponding to $10^8$ observed gamma-ray events. 
The vast majority of this simulated data consists of noise, but $100000$ events arise from the annihilation of scalar singlet dark matter with $m_\chi=\unit[1]{TeV}$.
We recover the dark matter mass within a 95\% credible interval of $m_\chi \sim \unit[0.96-1.07]{TeV}$.
Meanwhile, the velocity averaged cross section is constrained to $\langle\sigma v\rangle \sim \unit[1.4-2.1\times10^{-25}]{cm^3 s^{-1}}$ (95\% credibility).
This is equivalent to measuring the number of dark-matter annihilation events to be $N_S \sim 1.1_{-0.2}^{+0.2} \times 10^5$.
The no-signal hypothesis $\langle \sigma v \rangle=0$ is ruled out with about $5\sigma$ credibility. 
We discuss how \GammaBayes can be extended to include more sophisticated signal and background models and the computational challenges that must be addressed to facilitate these upgrades. 
The source code is publicly available \href{https://github.com/lpin0002/GammaBayes}{here}.}

\maketitle

\tableofcontents

\section{Introduction}
For decades there has been mounting evidence that cold, non-baryonic dark matter makes up a majority of the matter of the Universe 
\cite{Simon_2019, Allen_2011, 1993MNRAS.262.1023W}. 
Most of the evidence for dark matter is gravitational in nature, such as of anomalous galaxy rotation curves \cite{Zwicky1933, M33extended_velocity_curves} and gravitational lensing observations such as with the Bullet Cluster \cite{Bullet_Cluster__DM_Evidence}. 
While these measurements, together with observations of the CMB, show that dark matter makes up 85\% of the matter in the Universe, they do not tell us the dark-matter mass $m_\chi$ or how it interacts with itself and other standard-model particles \cite{Arcadi2017, Bertone2016}.
If dark matter interacts with standard-model particles, it may be possible to measure these quantities \cite{Slatyer2018, Drees2018}.
In particular, the annihilation of dark matter can produce high-energy photons, from regions of high dark-matter density such as the Galactic Center.
High-energy gamma-ray telescopes such as the  Cherenkov Telescope Array (CTA) \cite{CTAConsortium2017} therefore can provide a powerful indirect probe of dark matter particles with masses in the $m_\chi \approx \unit[10]{GeV}-\unit[10]{TeV}$ range \cite{Morselli2023, CTA2021}.

The CTA is an imaging atmospheric Cherenkov telescope array \cite{CTAConsortium2017}.
It works by capturing images of the flashes of Cherenkov radiation emitted from extensive air showers in the atmosphere, initiated by relativistic cosmic rays \cite{mazin2019cherenkov}.
The CTA, which is currently under construction, will eventually consist of two separate arrays: one in the Northern Hemisphere in La Palma, Spain near the MAGIC telescope, and another in the Southern Hemisphere in Paranal, Chile near the European Southern Observatory \cite{hofmann2023cherenkov}. 
The Southern site is well-positioned to observe the Galactic Center, which is a promising target for indirect dark matter searches~\cite{Morselli2023}.
The Galactic Center provides a relatively nearby and deep gravitational potential well which, as structure formation simulations indicate, contains a substantial amount of dark matter.
In this high-density environment, dark matter may self-interact, annihilating to particles that eventually produce gamma rays~\cite{Tulin_2018}. 
The Galactic Center, however, is also rich in astrophysical gamma-ray sources. 
The primary backgrounds for dark matter measurements include diffuse interstellar emission and localised gamma-ray sources from active galactic nuclei, pulsar wind nebulae, supernova remnants, and pulsars \cite{RowellBook}.
Weak, as-of-yet unresolved point sources are likely to be difficult to separate from a dark matter signal \cite{CTA2021}.
The challenge of disentangling a dark matter signal from astrophysical sources makes it difficult to confidently attribute gamma rays from the Galactic Center to dark matter annihilation \cite{Cholis2021, Slatyer2018}. 

Therefore, it is important to design dark matter detection pipelines that carefully model the distribution of signal and background in both energy and sky location.
In this work we present a Bayesian analysis pipeline to help address the challenge of detecting a dark matter signal from the Galactic Center with the CTA. 
The pipeline, called \texttt{GammaBayes}, is open source,\footnote{The code is available for download from \url{https://github.com/lpin0002/GammaBayes}.} and is also on the \texttt{Python} package index (\href{https://pypi.org/project/GammaBayes/}{\texttt{PyPI}}). 

The remainder of this paper is structured as follows. 
In section \ref{sec:Framework} we introduce the Bayesian formalism that underpins \texttt{GammaBayes}.
The instrument response functions (IRFs), accessed using \texttt{Gammapy}, are cast as likelihood functions.
The emitted energy spectra and sky position distribution expected for signal and background are cast as priors.
We largely follow the formalism from ref.~\cite{Mangipudi2021}, though, with more sophisticated modelling and new notation chosen to better adhere to other CTA literature.\footnote{For example, our analysis takes into account the energy-dependent effective area of the CTA, an important ingredient, which was omitted from ref.~\cite{Mangipudi2021}.}
In section \ref{sec:application} we apply \texttt{GammaBayes} to simulated data and demonstrate that we can correctly recover the dark-matter mass $m_\chi$ and velocity weighted annihilation cross section $\langle \sigma v \rangle$ for a scalar-singlet dark-matter model.
We estimate the sensitivity of the pipeline for various values of $m_\chi$ given the scalar singlet model.
In section \ref{sec:model_comparison} we then introduce the $\mathbb{Z}_5$ scalar dark matter model, and perform model comparison with this model and the $\mathbb{Z}_2$ scalar singlet dark matter model. 
In section \ref{sec:discussion} we discuss the limitations of the current version of \texttt{GammaBayes} and our plans for future improvements. 

\section{Formalism}\label{sec:Framework}
\subsection{CTA Data} 
Raw CTA data consists of images of particle showers measured by an array of optical telescopes.
Each candidate gamma-ray detection is referred to as an event.
The raw data for each event is pre-processed in order to obtain estimators for the reconstructed sky position $\mathbf{\Omega}_r$ and energy $E_r$.
These estimators are in contrast to the true sky position and the true energy $(\mathbf{\Omega}, E)$ (no subscript $r$).
Our starting point is this processed data.
We denote the data for event $i$ as $\mathbf{d}^i=(\mathbf{\Omega}^i_r, E^i_r)$.
The sky position includes two numbers: the Galactic latitude $b$ and the Galactic longitude $l$---the Galactic Center is situated at $(l,b)=(0,0)$.
For the sake of simplicity, our simulated data is generated assuming that the CTA is pointed directly at the Galactic Center so that the field-of-view (FOV) coordinates \cite{CTA2021, HESSGalacticPlaneSurvey2018} are identical to the Galactic coordinates.

\subsection{Likelihood}
\begin{figure*}[t]
    \centering\includegraphics[width=\textwidth]{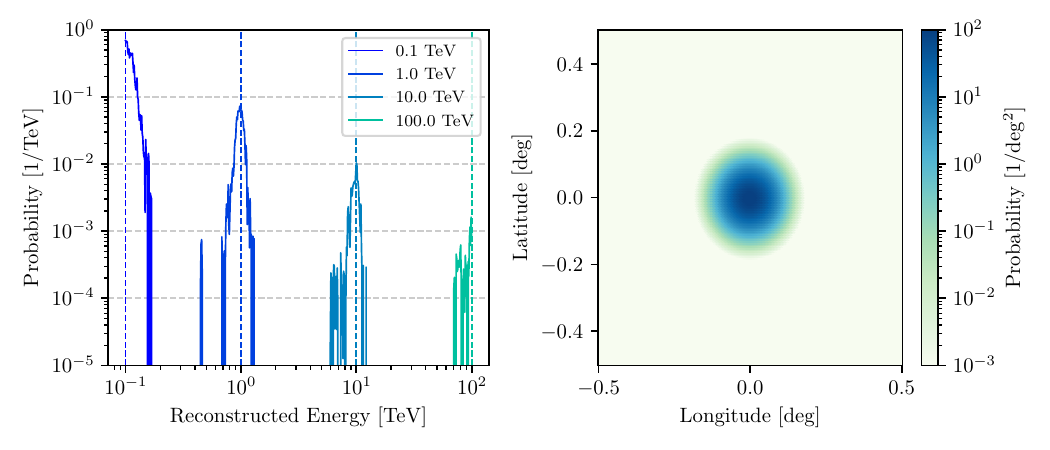}
    \caption{The two components of the CTA likelihood. (a) The likelihood for reconstructed energy; different colours correspond to different examples of true energy values. 
    The true sky location is (0,0). (b) The likelihood for reconstructed sky position given a true energy value of $\unit[1]{TeV}$ and true sky position of $(0,0)$.}
    \label{fig:edisppsf}
\end{figure*}

The likelihood function is a model of the measurement, quantifying the probability of observing $\mathbf{d}^i=(\mathbf{\Omega}^i_r, E^i_r)$ conditioned on the true values $(\mathbf{\Omega}^i, E^i)$.
The likelihood factorizes into two components: one for reconstructed energy and one for reconstructed sky position,
\begin{align}
    \mathcal{L} (\mathbf{d}^i|\mathbf{\Omega}^i, E^i) = & 
    \mathcal{L} (\mathbf{\Omega}_r^i, E_r^i |\mathbf{\Omega}^i, E^i) \nonumber\\
    = & \mathcal{L} (E_r^i|\mathbf{\Omega}^i, E^i)\mathcal{L}(\mathbf{\Omega}^i_r|\mathbf{\Omega}^i, E^i) .
\end{align}
Within the gamma-ray astronomy field these two components are referred to as the \textit{instrument response functions}. 
The likelihood of the reconstructed energy $E_r$, referred to as the \emph{energy dispersion}, is the probability density of the reconstructed energy given some true energy $E$ and true sky position $\mathbf{\Omega}$. 
The likelihood of the reconstructed sky position $\mathbf{\Omega}_r$, referred to as the \emph{point spread function}, is the probability density of the reconstructed sky position, given some true sky position $\mathbf{\Omega}$ and some true energy $E$. 
Example plots of the latest energy dispersion and point spread functions for the CTA are shown in the left and right panels of figure~\ref{fig:edisppsf}, respectively. 
We use the \texttt{prod5}, version-0.1 instrumental response functions available on \href{https://zenodo.org/record/5499840}{Zenodo} \cite{prod5irfsdataset}.\footnote{If in the future the instrument response functions are given as a single function, or single likelihood, the \texttt{GammaBayes} pipeline is capable of using them.}

\subsection{Signal prior}
The particle and astrophysics assumptions on dark matter are captured by a prior distribution of dark-matter gamma-ray annihilation products in energy and sky position:
\begin{align}
    \pi(\mathbf{\Omega}, E | \boldsymbol{\theta}, \mathcal{S}) .
\end{align}
It is a \textit{conditional prior}, the shape of which depends on the dark-matter parameters, $\boldsymbol{\theta}$, such as mass $m_\chi$, which in this context is a \textit{hyper-parameter}.
In principle, this prior can depend on additional hyper-parameters as well such as particle coupling constants. Additionally parameters describing the dark matter mass distribution about the Galactic Center could be included. However, to rigorously perform this analysis would lie outside the scope of this paper.
For the sake of simplicity, however, our prior from now on will only depend on $m_\chi$.
The $\mathcal{S}$ denotes that this prior describes a specific signal model.

\begin{figure}[t]
    \centering
    \begin{subfigure}[b]{0.49\textwidth}
        \centering
        \includegraphics[width=\textwidth]{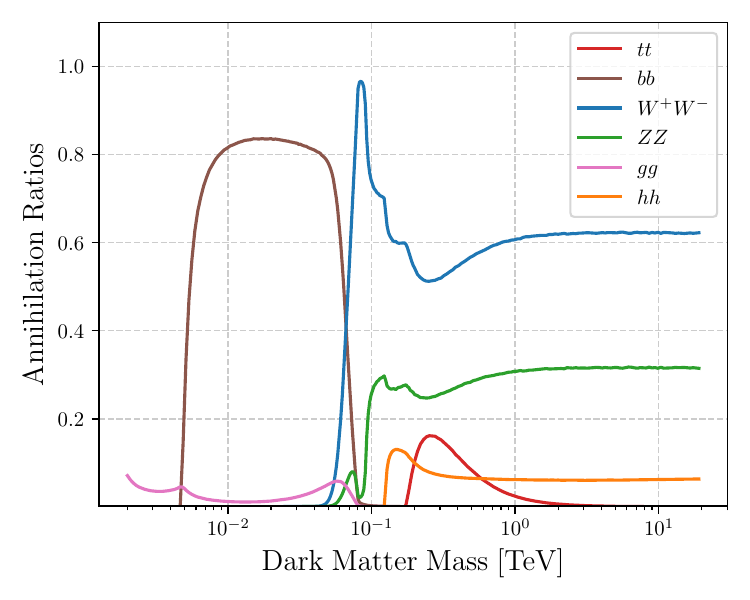}
        \label{fig:DM_branchingfractions}
    \end{subfigure}
    \begin{subfigure}[b]{0.49\textwidth}
        \centering
        \includegraphics[width=\textwidth]{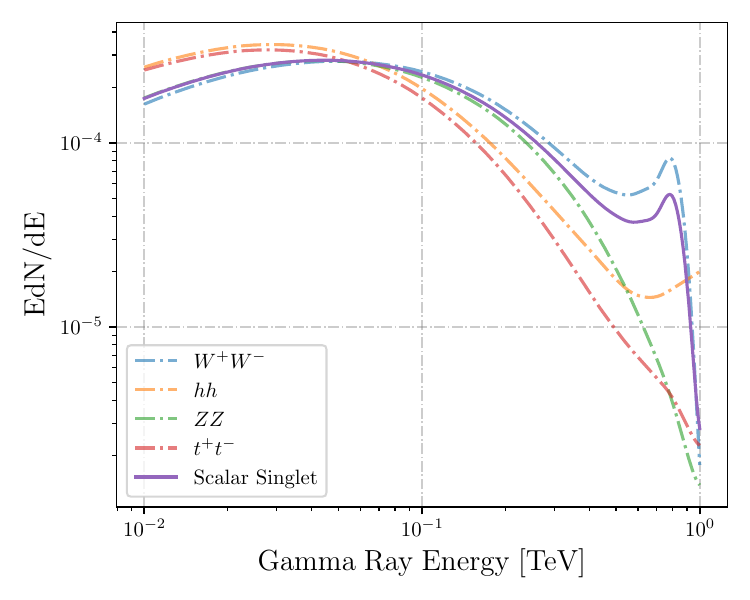}
        \label{fig:example_DM_spectra}
    \end{subfigure}
    \caption{(a) The non-negligible dark matter annihilation fractions versus dark-matter mass for the scalar singlet model; we fix $\lambda_{hS}=0.1$. 
    When  $m_\chi \gtrsim \unit[1]{TeV}$, the final products are consistently the $W$, $Z$ and Higgs bosons.
    (b) Example gamma-ray energy spectra for the scalar singlet model with the single channel contributions from the $W$, $Z$ and Higgs boson, which are the main final states for high masses. 
    There is a small resonance for dark matter masses higher than the $W$ boson, before a cut-off at $E = m_\chi$.}
    \label{fig:DM_Spectra_Plots}
\end{figure}

The energy component of the prior is proportional to the dark matter gamma ray annihilation spectrum which depends on the dark matter model presumed. For demonstration, we use the $\mathbb{Z}_2$ scalar singlet dark matter model, though we explain below how to adapt our framework to an arbitrary dark matter model.
The $\mathbb{Z}_2$ scalar singlet model is one of the simplest particle dark matter models. 
It adds a single massive scalar field to the standard model denoted $S$.\footnote{The scalar field $S$ is not to be confused with $\mathcal{S}$, which denotes the signal model.}  
The $\mathbb{Z}_2$ invariant terms of the Lagrangian for this model are given by \cite{GAMBIT2017_ScalarSinglet}.\footnote{Here, we use $\mathfrak{L}$ to denote the Lagrangian density in order to avoid confusing it with the likelihood function ${\cal L}$.}
\begin{align}
    \mathfrak{L} = \frac{1}{2}\mu_S S^2 + \frac{1}{2}\lambda_{hS}S^2|H|^2 + \frac{1}{4}\lambda_S S^4 +\frac{1}{2}\partial_\mu S \partial^\mu S .
\end{align}
These terms describe the bare $S$ mass, the Higgs-portal coupling, the $S$ quartic self-coupling, and the $S$ kinetic term respectively\footnote{Here and throughout the paper, we work in units of $c=1$.}. 
Assuming $S$ does not obtain a vacuum expectation value, the model contains only three free parameters: $\mu_S^2$, $\lambda_{hS}$ and $\lambda_S$. 
The tree-level mass of the singlet particle is
\begin{align}
    m_S = \sqrt{\mu_S^2+\frac{1}{2}\lambda_{hS}v_0^2},
\end{align}
where $v_0=\unit[246]{GeV}$ is the vacuum expectation value of the Higgs field. 
For indirect search methods we focus on the mass and Higgs-portal coupling as they predominately determine the properties of the gamma-ray spectrum produced. 
The scalar singlet dark-matter annihilation gamma-ray spectrum is characterized by a resonant bump followed by a hard cut-off at $E=m_\chi$; see figure~\ref{fig:DM_Spectra_Plots}b, which shows the spectrum for a scalar singlet model.
In our fits, we set $\lambda_{hS}$ to 0.1 and vary the mass, as variation of $\lambda_{hS}$ changes the overall shape of the spectra negligibly and essentially contributes a normalisation constant that is removed when the spectra is normalised to be used as a probability density. 
\color{black}{Using the code that reproduces the results shown in \cite{dimauro2023_SS_ratios} (found \href{https://github.com/dimauromattia/SingletScalar_DM/tree/main}{here})}\color{black}, we generate annihilation fractions of $S$ into various Standard Model final states for a range of masses; see~figure~\ref{fig:DM_Spectra_Plots}.  
It is at this stage of the framework that one can introduce a different dark matter model by plugging in a model's annihilation ratios\footnote{In searches assuming dark matter decay, a similar value is referred to as the `branching fraction' hence the use of $B_f$ within the formulae.}, to construct the dark matter spectra.
\color{black}{Using electroweak tabulated values calculated by the Poor Particle Physicists Cookbook \cite{PPPC_ref_1, PPPC_ref_2}}\color{black}, we interpolate the single-channel spectra and combine them using the annihilation ratios in order to produce energy spectra for the scalar singlet dark matter. 

An example spectrum for the scalar singlet model, created for a scalar mass $m_\chi = \unit[1]{TeV}$, is shown in figure~\ref{fig:DM_Spectra_Plots}. 
The prior for gamma-ray energy from dark matter annihilation is proportional to this spectrum.

\begin{figure}
    \centering
    \includegraphics[width=0.75\textwidth]{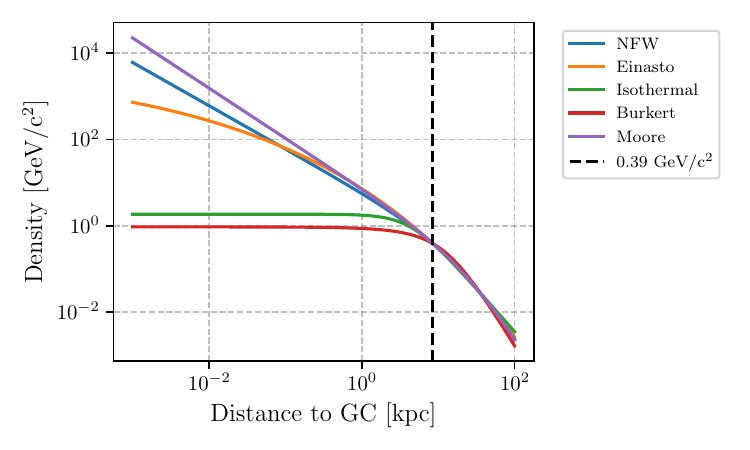}
    \caption{Various models for the dark matter density distribution within our galaxy. 
    The NFW, Einasto and Moore profiles continue to increase for small radii; they are examples of \emph{cuspy} profiles. 
    On the other hand, the Isothermal and Burkert distributions flatten off, and are thus examples of \emph{cored} profiles. 
    The vertical line represents the approximate local dark matter density of $\unit[0.39]{GeV/cm^3}$ \cite{Catena2009}.} 
    \label{fig:DM_Densities}
\end{figure}

The second component to the dark matter prior is its sky position distribution which is expected to be strongly peaked toward the Galactic Center; see figure~\ref{fig:DM_Densities}.
We use the differential $J$-factor based on the Einasto profile \cite{EinastoRef} defined as
\begin{align}\label{eq:diffJ}
    \frac{dJ}{d\mathbf{\Omega}} = \int_{\text{l.o.s}} d\ell(\mathbf{\Omega}) \rho_\chi^2(\mathbf{r}).
\end{align}
Here l.o.s refers to the ``line of sight'' between the observer and the dark matter annihilation, the axis of which is $\ell(\mathbf{\Omega})$.
Meanwhile, $\rho_\chi(\mathbf{r})$ is the dark matter density at position $\mathbf{r}$ relative to the Galactic Center.
We take the distance to the Galactic Center to be $\unit[8.5]{kpc}$ and the local dark matter density at the Sun to be $\unit[0.39]{GeV/cm^3}$ \cite{Catena2009}; see figure~\ref{fig:DM_Densities}. 
We use the CTA science tools package \texttt{Gammapy} \cite{Gammapy1, Gammapy2} to calculate these values.
The Einasto profile is an example of a \emph{cuspy} profile which has a large increase in the mass density near the Galactic Center as opposed to a \emph{cored} profile such as the Burkert profile \cite{BurkertRef}, which exhibits a near-flat density distribution close to the Galactic Center. 

\subsection{Background Prior}
The background prior is denoted:
\begin{align}
    \pi(\mathbf{\mathbf{\Omega}}, E | \mathcal{B} ) .
\end{align}
It describes the distribution of all gamma-ray-like events, which are not due to dark-matter annihilation.
In principle, the shape of the background prior can depend on hyper-parameters like the signal prior does.
For example, one can imagine introducing parameters describing the shape of the astrophysical gamma-ray flux from the Galactic Center.
However, for illustrative purposes, we assume the background is known precisely, and so there are no background hyper-parameters.

\begin{figure}
    \centering
    \includegraphics[width=0.75\textwidth]{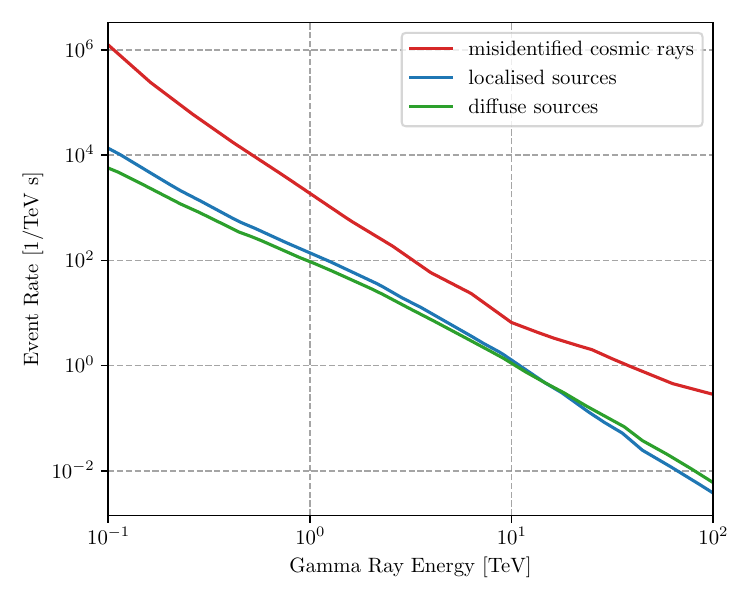}
    \caption{
    Gamma-ray energy spectra contributing to the background prior. 
    The main contribution is from misidentified cosmic rays.
    }
    \label{fig:bkg_spectra}
\end{figure}

\begin{figure}
    \centering
    \includegraphics[width=0.75\textwidth]{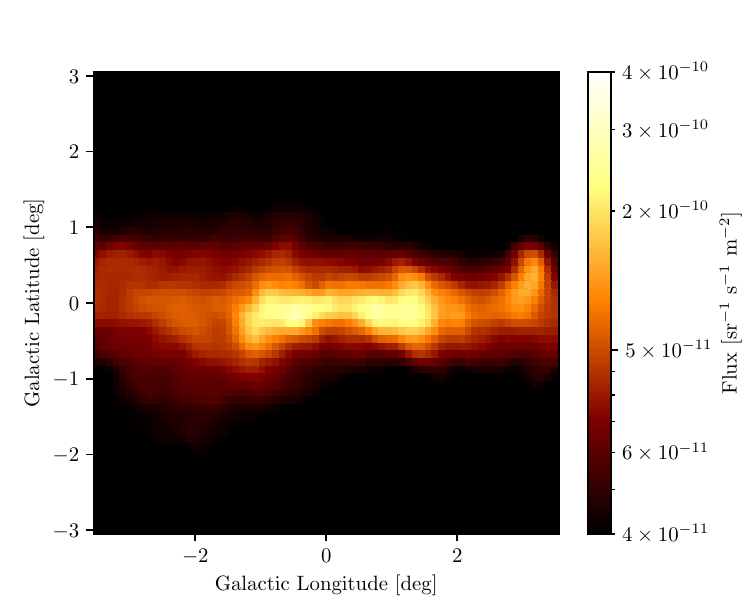}
    \caption{Sky map of the diffuse Fermi-LAT background, integrated over energy.}
    \label{fig:bkg_morphology}
\end{figure}

Our background model consists of three components\footnote{The models are all accessed using the python package \texttt{Gammapy}}: diffuse flux from astrophysical gamma-ray sources clustered about the Galactic plane, astrophysical point sources, and misidentified cosmic rays.
For the diffuse component, the sky position distribution is based on the Fermi-LAT Galactic diffuse model used for the fourth Fermi Large Area Telescope source catalog analysis.\footnote{The model is contained in the \texttt{gll\_iem\_v06\_gc.fits.gz} file in the \texttt{Gammapy} package.}
The energy distribution for the diffuse model is then the same as the power-law fit in ref. \cite{Gaggero_2017}, based on observations within the \emph{pacman} region around the Galactic Center. 
Graphical representations of the background model are shown in figure~\ref{fig:bkg_spectra} and figure~\ref{fig:bkg_morphology}.

According to the HESS catalogue there are seven sources within $5^\circ$ of the Galactic Center; we include these in our background model.\footnote{The identified sources are J1741-302 (UNID), J1745-290 (Galactic Center/UNID), J1745-303 (SNR/molecular cloud), J1746-285 (UNID), J1746-308 (UNID), J1747-248 (UNID), and J1747-281 (PWN).} 
There are likely many unidentified point sources near the Galactic Center due to its complex source morphology. 
The only confidently known source in this region is J1747-281, which is associated with a pulsar wind nebula that primarily emits gamma rays through an inverse Compton scattering process initiated by electrons accelerated by a termination shock front \cite{Hillas_1998, High_and_Low_Energy_PWNe_spectra, RowellBook}.
We do not mask these sources, but opt instead to model them.
(Gamma rays pointing toward these point sources do not provide much support for dark-matter annihilation because they can be explained with the background model.)
The energy-integrated spectrum for this component as seen by the CTA is shown in figure~\ref{fig:bkg_spectra}.

\begin{figure}
    \centering
    \includegraphics[width=0.75\textwidth, trim={0 0.5cm 0 0}]{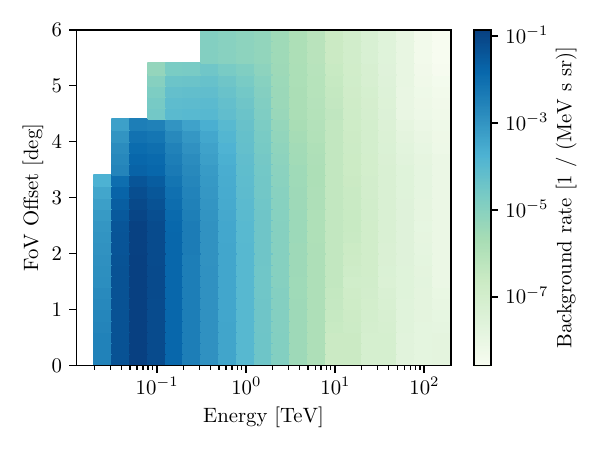}
    \caption{The \texttt{prod5} CTA cosmic-ray misidentification rate---an ingredient in the background prior. For large field-of-view offsets the rate significantly decreases because the atmosphere is effectively thicker for larger offset angles and so events are less likely to be detected in general.}
    \label{fig:IRFBKG}
\end{figure}

The final component of our background model is the \texttt{prod5} version of misidentified cosmic-ray background as shown in figure~\ref{fig:IRFBKG}. 
This component represents the background of charged cosmic rays that pass the CTAs selection cuts. 
This background is described in the frame of the telescope; the background varies according to the offset from the center of the pointing direction.

\subsection{Effective Area}\label{sec:EffectiveArea}
\begin{figure}
    \centering
    \includegraphics[width=0.75\textwidth, trim={0 0.5cm 0 0}]{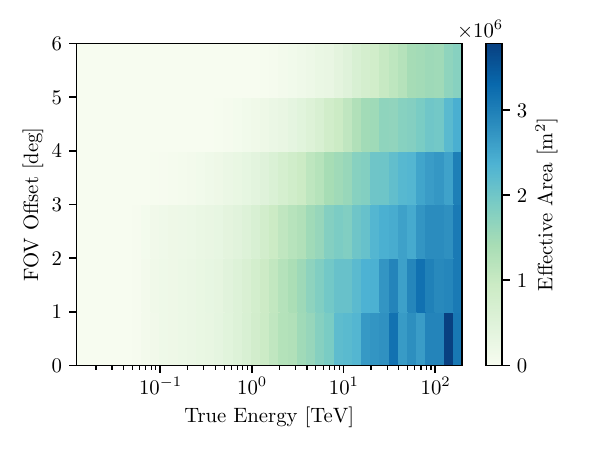}
    \caption{Effective area of the CTA South site for a zenith angles between 0 and 20 degrees as a function of true energy and offset between the reconstructed sky position and true sky position.}
    \label{fig:aeff}
\end{figure}

The signal and background models defined in the previous two subsections describe the distribution of gamma rays \textit{emitted} by various sources in and around the Galactic Center.
However, the CTA observes a different distribution due to its effective area $A_\text{eff}$, which depends on both gamma-ray energy and also on the offset from the center of the field of view; see figure~\ref{fig:aeff}.
Due to this dependence, the CTA is more sensitive to high-energy gamma rays than it is to low-energy gamma rays.
The \textit{observed} distribution is related to the emitted distribution like so:\footnote{The misidentification background is in the frame of the telescope so it is already the observed distribution.}
\begin{align}
    \pi(\mathbf{\Omega}, E)_\text{observed} \propto A(E, \mathbf{\Omega}) \,
    \pi(\mathbf{\Omega}, E)_\text{emitted} .
\end{align}
In the discussion that follows, all of the priors we refer to are priors for \textit{observed} gamma rays, which take into account the effective area of the CTA South site for a zenith angles between 0 and 20 degrees.

\subsection{Combined likelihood}
We now have the ingredients to construct a likelihood function for the entire data set given the dark-matter parameters.
First, we consider a single event with data $\mathbf{d}_i$.
Since we do not know whether any given event is due to signal or background, the likelihood of $\mathbf{d}_i$ can be expressed as a  mixture model of signal $\mathcal{S}$ and background $\mathcal{B}$:
\begin{align}
    \mathcal{L}(\mathbf{d}^i|m_\chi, \xi) = \xi \, \mathcal{L}(\mathbf{d}^i|m_\chi, \mathcal{S}) + (1-\xi)\, \mathcal{L}(\mathbf{d}^i|\mathcal{B}) .
    \label{eqn:single_event_mixture_likelihood}
\end{align}
Here, $\xi$ represents the probability that the event is drawn from the signal distribution. 
Meanwhile,
\begin{align}\label{eq:ZSZN}
\mathcal{L}(\mathbf{d}^i|m_\chi, \mathcal{S}) = & \int d\mathbf{\Omega}^i \int dE^i \mathcal{L}(d^i|\mathbf{\Omega}^i, E^i) \pi(\mathbf{\Omega}^i, E^i|m_\chi, \mathcal{S}) \\
\mathcal{L}(\mathbf{d}^i|\mathcal{B}) = & \int d\mathbf{\Omega}^i \int dE^i \mathcal{L}(\mathbf{d}^i|\mathbf{\Omega}^i, E^i) \pi(\mathbf{\Omega}^i, E^i|\mathcal{B}),
\label{eqn:bkg_marginal_likelihood}
\end{align}
are \textit{marginal likelihoods} for the signal and background models.

Since the events are uncorrelated, the combined likelihood for all data $\vec{d} = \{\mathbf{d}^i\}$ is the product of $N$ single-event likelihoods:
\begin{align}
    \mathcal{L}(\vec{d}|m_\chi, \xi) = \prod^N_i \mathcal{L}(\mathbf{d}^i|m_\chi, \xi) .
\end{align}
In this context, one can interpret $\xi$ as the fraction of events that are drawn from the signal distribution $N_S$:
\begin{align}
    \xi = \frac{N_{\mathcal{S}}}{N} .
\end{align}
The $\xi$ parameter is sometimes referred to as the ``mixing fraction'' or a ``duty cycle''.
In the next subsection we describe how $\xi$ relates to dark-matter parameters.

We use the combined likelihood to calculate the posterior probability distribution for parameters $\xi, m_\chi$ given the data:
\begin{align}\label{eq:posterior}
    p(\xi, m_\chi | \vec{d}) = \frac{\mathcal{L}(\vec{d}|m_\chi, \xi)\pi(m_\chi)\pi(\xi)}{\mathcal{Z}(\vec{d})} .
\end{align}
Here, $\pi(m_\chi)$ is the prior on the dark-matter mass, which we take to be log-uniform on the interval $(\unit[100]{GeV},\unit[100]{TeV})$.
Meanwhile, $\pi(\xi)$ is the prior on mixing fraction, which we take to be uniform on the interval $(0,1)$. 
The Bayesian evidence $\mathcal{Z}(\vec{d})$ is the fully marginalised likelihood. 
It appears in eq.~\ref{eq:posterior} as a normalization factor, but we discuss below how it can be used for model selection; see eq.~\ref{eq:evidence}.

\subsection{Dark matter parameters}\label{sec:DM_Parameters}
We relate $\xi$ to the dark matter parameters through the relation,
\begin{align}\label{eq:sigmav_mass_and_chi_relation}
    \xi &= \frac{T_\text{obs}}{N} \int_{\Delta \mathbf{\Omega}} d\mathbf{\Omega} \int dE A_\text{eff}(E, \mathbf{\Omega}) \frac{d\Phi}{d\mathbf{\Omega} dE} .
\end{align}
Here, $T_\text{obs}$ is the observation time and $d\Phi/d\mathbf{\Omega} dE$ is the differential flux of gamma rays produced by dark matter annihilation per unit energy and solid angle given by,
\begin{align}
   \frac{d\Phi}{d\mathbf{\Omega} dE} &= \frac{1}{4\pi}\frac{dJ}{d\mathbf{\Omega}} \frac{\langle \sigma v\rangle_\text{ann}}{2S_\chi m_\chi^2} \sum_f B_f \frac{dN^f}{dE}.
\end{align}
Here, $S_\chi$ is a symmetry factor representing whether the dark matter particle is its own antiparticle ($S_\chi=1$) or not ($S_\chi=2$).
The factor ${dJ}/{d\mathbf{\Omega}}$ is as defined in eq.~\ref{eq:diffJ}.
The summation is over all the possible dark matter annihilation final states $f$ including gauge bosons, quarks or leptons. 
The annihilation fraction $B_f$ is the ratio of the annihilation cross section for final state $f$ to the total annihilation cross section,
\begin{align}
    B_f = \frac{\sigma_f}{\sigma_\text{tot}}.
\end{align}
Here $\sigma_\text{tot}$ refers to the ``total'' annihilation cross-section defined as the sum of all the individual final state cross-sections and the distribution ${dN^f}/{dE}$ represents the differential energy flux of gamma rays emitted from these final states. 

Finally, $\langle \sigma v \rangle$ represents the thermally-averaged, velocity-weighted total annihilation cross section. We then assume the low-velocity limit which, for the scalar singlet model, allows us to approximate $\langle \sigma_f v \rangle \approx \sigma_f v$, so the equation becomes, 
\begin{align}
    \frac{d\Phi}{d\mathbf{\Omega} dE} &= \frac{1}{4\pi} \frac{dJ}{d\mathbf{\Omega}} \frac{1}{2S_\chi m_\chi^2} \sum_f \langle \sigma_f v \rangle \frac{dN^f}{dE} .
\end{align}
The essence of this subsection is that the mixing fraction $\xi$ and mass $m_\chi$ are mappable to $\langle \sigma v \rangle$, and so the posterior for $\xi$ and $m_\chi$ can be converted into a posterior on $\langle \sigma v \rangle$.

\subsection{Computer code}
We implement the analysis described above within a pip-installable \texttt{Python} package, called \texttt{GammaBayes}. 
It is built in such a way that it is easy-to-use and modular so that a user can change the various models and assumptions discussed above. 
For example one can change the dark matter model by changing the input annihilation fractions (differential cross-sections).
Or, if the CTA point spread function model is updated, one can swap in a new IRF.
Core functions use the \texttt{Gammapy} \texttt{Python} package such as the calculation of the differential $J$-factors for the Einasto dark matter mass distribution. 
\texttt{GammaBayes}, is open source, available at \url{https://github.com/lpin0002/GammaBayes}.

\section{Demonstration}\label{sec:application}
In this section we apply \texttt{GammaBayes} to simulated data in order to demonstrate the detection of a signal arising from the annihilation of scalar singlet dark matter.
We simulate $\unit[525]{hours}$ of data, corresponding to approximately $10^8$ gamma-ray events assuming that the backgrounds are stable, or that transient sources are excluded. 
More realistic analyses would require specific IRFs provided for each individual observation run. This cannot be realistically done until actual data has been collected. 
However, if relevant IRFs are chosen, it should not cause any significant changes to our results.

The vast majority of these simulated events are drawn from the background distribution.
However, there are $10^5$ events drawn from the signal distribution (corresponding to a mixing fraction $\xi = 10^{-3}$) with dark-matter mass $m_\chi = \unit[1]{TeV}$.

\begin{figure*}
    \centering
    \includegraphics[width=0.9\textwidth]{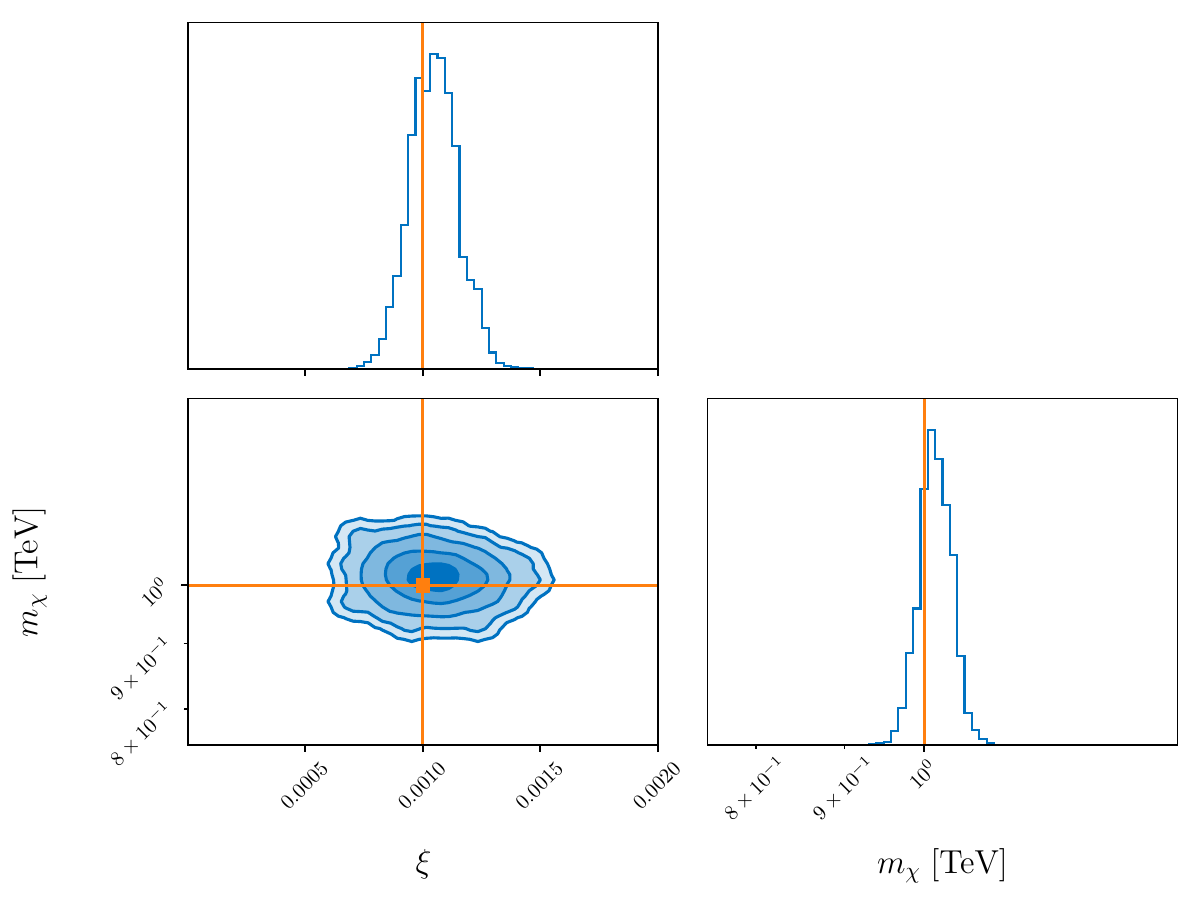}
    \caption{
    A corner plot showing the posterior distribution for $\unit[525]{hours}$ of simulated CTA Galactic Center survey data corresponding to $10^8$ gamma-ray events.
    The data contains 100,000 gamma rays from $m_\chi = \unit[1]{TeV}$ dark-matter annihilation corresponding to a mixing fraction $\xi=10^{-3}$.
    The values of these parameters are indicated with the orange lines.
    The bottom-left panel shows the credible intervals from one to five sigma.
    The dark matter mass $m_\chi$ is constrained to $\unit[1.02^{+0.04}_{-0.06}]{TeV}$ and the mixing fraction is constrained to $1.1^{+0.2}_{-0.2}\times10^{-3}$ (95\% credibility).
    The null hypothesis that no dark-matter signal is present in the data $\xi=0$ is excluded with five sigma credibility.
    }
    \label{fig:example_posterior_result}
\end{figure*}

\begin{figure}
    \centering
    \includegraphics[width=0.75\textwidth]{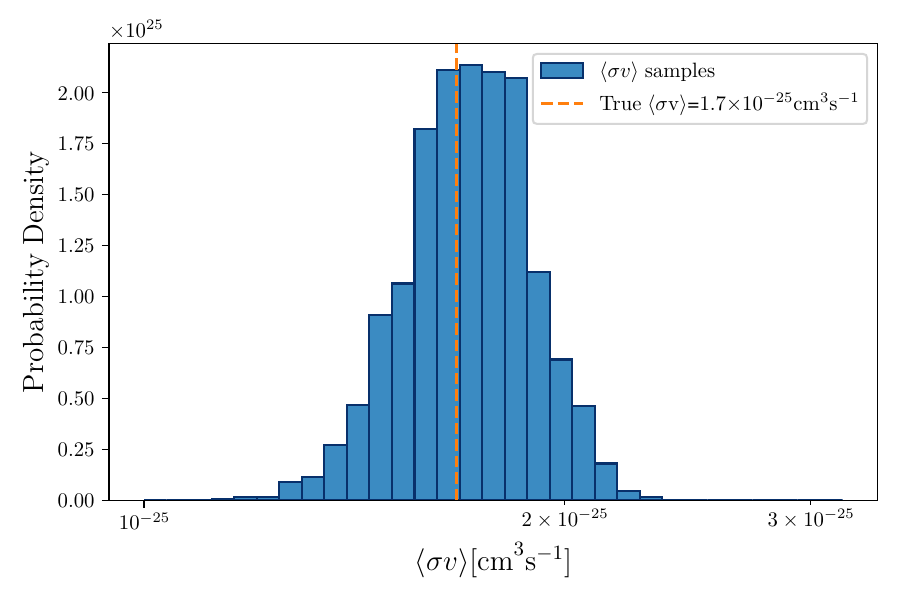}
    \caption{The posterior for $\langle \sigma v \rangle$. The orange dashed line represents the true value of $\langle \sigma v\rangle$.
    }
    \label{fig:sigmav_posterior}
\end{figure}

For each event, we simulate the reconstruction of the gamma-ray energy on the interval $(\unit[100]{GeV}, \unit[300]{TeV})$ approximately corresponding to the expected observable energy range of the CTA \cite{hofmann2023cherenkov}. 
We simulate the reconstruction of the sky position in a box centered on the Galactic Center and spanning $7^\circ$ along the axis of Galactic longitude and $6^\circ$ along the axis of Galactic latitude.\footnote{The size of this box is arbitrary. It should be large enough to include the region of parameter space where the signal prior is large. It could be made larger, but the gain in sensitivity would be marginal.}
Applying \texttt{GammaBayes} to this simulated data yields a joint posterior for mixing fraction $\xi$ and $m_\chi$, which is shown in  figure~\ref{fig:example_posterior_result}.
The true value of each parameter is indicated using orange lines.
The levels indicate the one to five sigma credibility intervals.
The dark matter mass is constrained to $\unit[1.02^{+0.04}_{-0.06}]{TeV}$ 
while the mixing fraction is constrained to $1.1^{+0.2}_{-0.2}\times10^{-3}$ (both 95\% credibility).
The mixing fraction excludes $\xi=0$ with five-sigma credibility ($\xi=0$ falls outside the highest probability density interval that includes all but $3\times10^{-5}$ of the probability).
Figure~\ref{fig:example_posterior_result} is therefore an illustration of a $5\sigma$ dark-matter detection.
This posterior can also be converted into a posterior on $\langle \sigma v \rangle$ as discussed in section~\ref{sec:DM_Parameters}, this result is shown in figure~\ref{fig:sigmav_posterior}.

\begin{figure}
    \centering
    \includegraphics[width=0.75\textwidth]{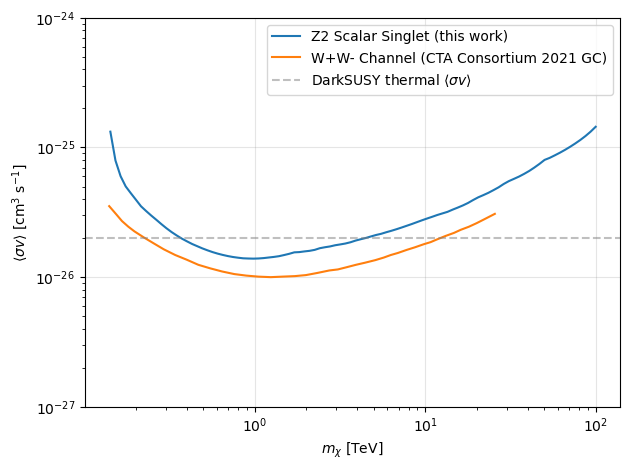}
    \caption{
    The sensitivity of the CTA to dark-matter annihilation from the annihilation of scalar single particles \color{black}{using annihilation ratios supplied by the \texttt{darkSUSY} code package \cite{Bringmann_2018, P_Gondolo_2004}.}\color{black}
    In blue we plot the expected 95\% credibility upper limits as a function of $m_\chi$.
    In orange, meanwhile, is the expected frequentist upper limits from \cite{CTA2021}. 
    The two curves should be compared \textit{qualitatively} since the two analyses employ different signal and background models; see the main text for details.
    The horizontal grey dashed line represents the $\langle \sigma v \rangle$ value inferred from cosmological calculations for the current dark matter relic density.}
    \label{fig:sensitivity_plot}
\end{figure}

In the event that the posterior for mixing fraction is consistent with $\xi=0$, then the data provides no evidence for dark-matter annihilation, and so we set an upper limit on $\xi$.
In figure~\ref{fig:sensitivity_plot}, we plot in blue the average 95\% credibility upper limit on $\langle \sigma v \rangle$ as a function of the dark-matter mass using annihilation ratios supplied by the \texttt{darkSUSY} code package \cite{Bringmann_2018, P_Gondolo_2004}.
Here, we also use $\unit[525]{hrs}$ of the CTAs central Galactic Center survey with no dark matter events.
For comparison, we show in orange the expected sensitivity for the $WW$ channel estimated in ref.~\cite{CTA2021}.
We caution the reader: this is an apples-to-oranges comparison, useful only for order-of-magnitude checks and qualitative features.
The two analyses employ different background models, different instrument response functions, and different dark matter annihilation final states. Additionally, the blue curve is a Bayesian credible interval while the orange is a frequentist confidence interval.
With that important caveat in mind, the two curves are similarly shaped with similar minimum values of $\langle\sigma v \rangle$.\footnote{The fact that the two curves match closely around $m_\chi \approx \unit[1]{TeV}$ is likely a numerical coincidence, and indeed, the curves diverge for lower/higher values of dark-matter mass.}
The horizontal grey dashed line represents the $\langle \sigma v \rangle$ value inferred from cosmological calculations for the current dark matter (relic) density.
Thus, given our signal and background models, we forecast that the CTA is capable of seeing at least weak $\gtrsim 2\sigma$ evidence of dark matter annihilation in the range of $(\unit[400]{GeV}, \unit[5]{TeV})$.

\section{Model Selection}\label{sec:model_comparison}
\texttt{GammaBayes} can also be used to perform Bayesian model selection.
In addition to producing a posterior distribution, \texttt{GammaBayes} calculates the Bayesian evidence for the data given the user-specified model ${\cal M}$, for example,:
\begin{align}\label{eq:evidence}
    {\cal Z}(\vec{d}|{\cal M}) = \int  dm_\chi 
    \int d\xi \, 
    {\cal L}(\vec{d} | m_\chi, \xi) \,
    \pi(m_\chi | {\cal M}) \, \pi(\xi | {\cal M}) .
\end{align}
By taking the ratio of two evidence values (for two different models ${\cal M}_1$ and ${\cal M}_2$), we obtain a Bayes factor,
\begin{align}
    \text{BF}^2_1 = \frac{\mathcal{Z}(\vec{d}|\mathcal{M}_2)}{\mathcal{Z}(\vec{d}|\mathcal{M}_1)} ,
\end{align}
which can be used to see, which model better describes the data.
A large Bayes factor indicates one model better explains the data.
Sometimes a threshold of
\begin{align}
    \ln \left(\left|\text{BF}\right|\right) > 8 ,
\end{align}
is used for designating an overwhelmingly statistically significant preference \cite{jeffreys1961theory, Thrane2018}.
The Bayes factor compares not just the goodness of fit for each model, but also includes Occam factors, which penalize overly flexible models \cite{Loredo2012}. 
The Occam factor is a mathematical formulation of the maxim that the simplest explanation is most likely correct, all else being equal.

\color{black}{In order to demonstrate Bayesian model selection with \texttt{GammaBayes}, we test whether the data is better explained by the $\mathbb{Z}_2$ scalar singlet model described above or a $\mathbb{Z}_5$ dark matter model such as in Ref.~\cite{Z5BalangerRef}. 
The Lagrangian formulation for the $\mathbb{Z}_5$ model is given by,
\begin{align}
    \mathfrak{L} &= \mu_1^2|\phi_1|^2 + \lambda_{41}|\phi_1|^4 + \lambda_{S1}|H|^2|\phi_1|^2+\mu_2^2|\phi_2|^2+\lambda_{42}|\phi_2|^2|\lambda_{S2}|H|^2|\phi_2|^2 \nonumber\\&+ \lambda_{412}|\phi_1|^2|\phi_2|^2 + \frac{1}{2}[\mu_{S1}\phi_1^2\phi_2^* + \mu_{S2}\phi_2^2\phi_1 + \lambda_{31}\phi_1^3\phi_2 + \lambda_{32}\phi_1\phi_2^{*3} + \rm{H.c.}].
\end{align}
We extract the annihilation ratios from the package \verb|micrOMEGAs|, fixing all parameters except for one dark matter particle mass for this model to within the possible regions outlined in \cite{Z5BalangerRef} (trading $\mu_1$ and $\mu_2$ for the masses). 
The fixed values are given in Table~\ref{tbl:fixed_Z5_parameters} and example annihilation ratios and spectra (for particular parameter values) are shown in Fig.~\ref{fig:Z5_ratios}.
A more general model (that includes other sub-models as special cases) will always produce a better fit (a higher likelihood) than its sub-models.

\begin{table}[h!]
\centering
\begin{tabular}{|c c|}
    \hline
    Parameter &Value \\
    \hline
    $\lambda_{41}$ & 0.001 \\
    $\lambda_{S1}$ & 0.1 \\
    $\lambda_{42}$ & 0.001 \\
    $\lambda_{S2}$ & 0.1 \\
    $\lambda_{412}$ & 0.0 \\
    $\mu_{S1}$ & $\unit[500]{TeV}$ \\
    $\mu_{S2}$ & $\unit[0.0]{TeV}$ \\
    $\lambda_{31}$ & 0.0 \\
    $\lambda_{32}$ & 0.0 \\
    $m_{\chi2}$ & $\unit[0.7]{TeV}$\\
    \hline
\end{tabular}
\caption{ \color{black}{The fixed parameters used within the analysis involving the $\mathbb{Z}_5$ dark matter model.}\color{black}}
\label{tbl:fixed_Z5_parameters}
\end{table}

\begin{figure}[h!]
    \centering
    \begin{subfigure}[t]{0.48\textwidth}
        \centering
        \includegraphics[width=\textwidth]{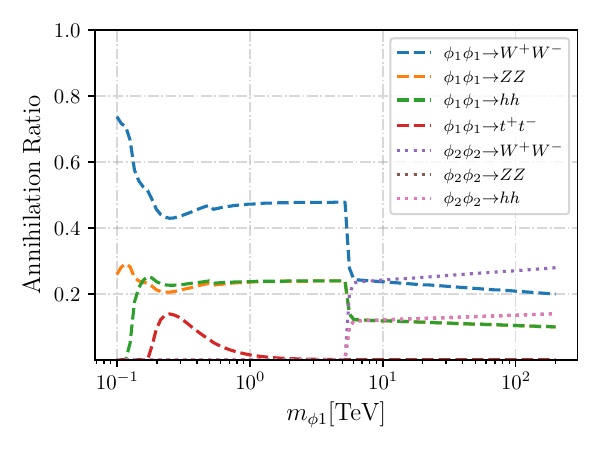}
    \end{subfigure}
    ~
    \begin{subfigure}[t]{0.48\textwidth}
        \centering
        \includegraphics[width=\textwidth]{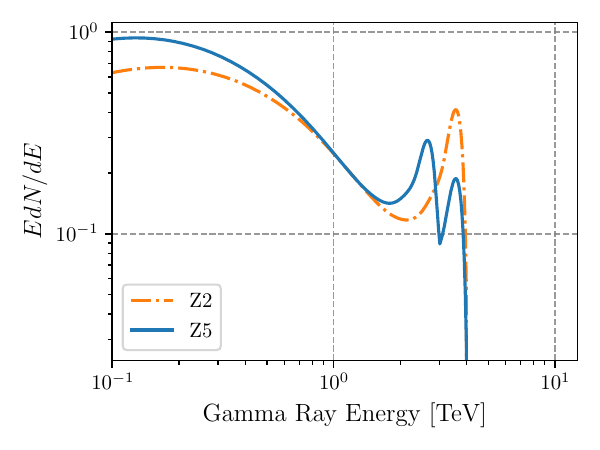}
    \end{subfigure}

    \caption{(a) \color{black}Spectrum of the {$\mathbb{Z}_5$ scalar dark matter annihilation ratios that contribute more than 5\% of the total for some value of $m_{\phi 1}$. We employ the parameters shown in Table \ref{tbl:fixed_Z5_parameters} with $m_{\phi2}=$\unit[7]{TeV} using micrOMEGAs 6.0 \cite{micrOMEGAs6p0}. (b) The subsequent normalised gamma ray spectrum for the $\mathbb{Z}_5$ dark matter model for $m_{\phi 1} =$ \unit[3]{TeV} and $m_{\phi 2} =$ \unit[4]{TeV} against the $\mathbb{Z}_2$ scalar singlet dark matter spectrum for a mass of \unit[4]{TeV} for comparison. The double peak in the $\mathbb{Z}_5$ dark matter spectrum comes from the annihilation of the two dark matter particle species.}\color{black}}
    \label{fig:Z5_ratios}
\end{figure}

Using our simulated data for the example detection, which includes $10^5$ gamma ray events from the $\mathbb{Z}_2$ scalar singlet dark matter model (signal fraction $\xi=10^{-3}$), we perform demonstrative Bayesian model selection using \texttt{GammaBayes} by computing $\ln{\cal Z}$ values for both models.
We find that the $\mathbb{Z}_2$ model is overwhelmingly preferred over the $\mathbb{Z}_5$ model with $ln \,\text{BF} = 16.8$.
Interestingly, if we were to obtain this Bayes factor with real CTA data, it would not imply that the $\mathbb{Z}_2$ is the correct model---only that it is strongly preferred over the $\mathbb{Z}_5$ model.
Subsequent particle physics experiments might be required to confirm the true nature of dark matter.}\color{black}


\section{Discussion and conclusions}\label{sec:discussion}
In our demonstration we assume that our signal and background models are adequately specified. 
In reality, however, these models are subject to systematic uncertainty. 
For example, the dark matter density profile is uncertain \cite{ZuriagaPuig2023}. 
While we employ the cuspy Einasto profile, the actual distribution of dark matter may be better described by the Burkert profile, which is nearly flat for small radii \cite{bouche_coredDM_startforming_galaxies}.
A misspecified model for the distribution of dark matter could yield unreliable inferences on $\xi$---Bayesian inferences are as reliable as the assumptions that underpin them.
The solution is to build flexible models that incorporate theoretical uncertainty, e.g., with hyper-parameters that allow for different plausible profile shapes.
By marginalising over these hyper-parameters, it is possible to include systematic uncertainty in the posterior for $m_
\chi, \xi$.
Future \texttt{GammaBayes} development will include the development of such flexible models.
\color{black}
We also demonstrate the ability to distinguish different dark matter models. 
\color{black}

Another priority is to improve the computational efficiency of the analysis in order to reduce the calculation run time. 
Currently, our analysis is carried out on a grid of the prior parameters (e.g. $m_\chi$).
However, this is inefficient because computer cycles are wasted exploring low-likelihood regions.
By employing a stochastic sampler to marginalise the nuisance parameters, we aim to significantly improve efficiency in addition to other improvements to the computation speed with the use of GPUs.


\acknowledgments{This work was performed on the OzSTAR national facility at Swinburne University of Technology. The OzSTAR program receives funding in part from the Astronomy National Collaborative Research Infrastructure Strategy (NCRIS) allocation provided by the Australian Government, and from the Victorian Higher Education State Investment Fund (VHESIF) provided by the Victorian Government.
E.T. is supported by ARC CE170100004, LE210100002, DP230103088, and CE230100016.
The research of C.B. is supported by ARC DP210101636, DP220100643, and LE210100015.}

\bibliography{refs}

\end{document}